\documentclass[pdflatex,sn-aps]{sn-jnl}

\usepackage{graphicx}%
\usepackage{multirow}%
\usepackage{amsmath,amssymb,amsfonts}%
\usepackage{amsthm}%
\usepackage{mathrsfs}%
\usepackage[title]{appendix}%
\usepackage{xcolor}%
\usepackage{textcomp}%
\usepackage{manyfoot}%
\usepackage{booktabs}%
\usepackage{algorithm}%
\usepackage{algorithmicx}%
\usepackage{ulem}
\usepackage{algpseudocode}%
\usepackage{listings}%
\usepackage[right]{lineno}

\theoremstyle{thmstyletwo}%

\theoremstyle{thmstylethree}%

\begin{document}

\title[Article Title]{Proximity-Induced Nodal Metal in an Extremely Underdoped CuO$_2$ Plane in Triple-Layer Cuprates}


\author*[1,2]{\fnm {Shin-ichiro} \sur{Ideta}}\email{idetas@hiroshima-u.ac.jp}
\author[3,4]{\fnm{Shintaro} \sur{Adachi}}
\author[5]{\fnm{Takashi} \sur{Noji}}
\author[3]{\fnm{Shunpei} \sur{Yamaguchi}}
\author[3]{\fnm{Nae} \sur{Sasaki}}
\author[6]{\fnm{Shigeyuki} \sur{Ishida}}
\author[6,9]{\fnm{Shin-ichi} \sur{Uchida}}
\author[8]{\fnm{Takenori} \sur{Fujii}}
\author[3]{\fnm{Takao} \sur{Watanabe}}
\author[7, 10, 11]{\fnm{Wen O.} \sur{Wang}}
\author[10]{\fnm{Brian} \sur{Moritz}}
\author[10, 12, 13]{\fnm{Thomas P.} \sur{Devereaux}}
\author[1]{\fnm{Masashi} \sur{Arita}}
\author[14]{\fnm{Chung-Yu} \sur{Mou}}
\author[15]{\fnm{Teppei} \sur{Yoshida}}
\author[2]{\fnm{Kiyohisa} \sur{Tanaka}}
\author[14]{\fnm{Ting-Kuo} \sur{Lee}}
\author*[9,14]{\fnm{Atsushi} \sur{Fujimori}}\email{fujimori@phys.s.u-tokyo.ac.jp}
\affil*[1]{\orgname{Research Institute for Synchrotron Radiation Science, Hiroshima University}, {\city{Higashi Hiroshima}, \postcode{7390046}, \country{Japan}}}
\affil[2]{\orgname{UVSOR-III Synchrotron, Institute for Molecular Science}, {\city{Okazaki}, \postcode{4448585}, \country{Japan}}}
\affil[3]{\orgdiv{Graduate School of Science and Technology}, \orgname{Hirosaki University}, {\city{ Hirosaki}, \postcode{036-8561}, \country{Japan}}}
\affil[4]{\orgdiv{Kyoto Univ Adv Sci, Nagamori Inst Actuators}, {\city{Kyoto}, \postcode{615-8577}, \country{Japan}}}
\affil[5]{\orgdiv{Graduate School of Engineering, Tohoku University}, \orgaddress{\city{Sendai}, \postcode{980-8561}, \country{Japan}}}
\affil[6]{\orgdiv{National Institute of Advanced Industrial Science and Technology}, \orgaddress{\city{Tsukuba}, \postcode{305-8568}, \country{Japan}}}
\affil[7]{\orgdiv{Department of Applied Physics, Stanford University, Stanford}, \orgaddress{\postcode{94305}, \state{CA}, \country{USA}}}
\affil[8]{\orgdiv{Cryogenic Research Center, The University of Tokyo}, \orgaddress{\city{Tokyo}, \postcode{113-0032}, \country{Japan}}}
\affil[9]{\orgdiv{Department of Physics, University of Tokyo}, \orgaddress{\city{Tokyo}, \postcode{113-0032}, \country{Japan}}}
\affil[10]{\orgdiv{Stanford Institute for Materials and Energy Sciences, SLAC National Accelerator Laboratory}, \orgaddress{\postcode{94025}, \state{CA}, \country{USA}}}
\affil[11]{\orgdiv{Kavli Institute for Theoretical Physics, University of California, Santa Barbara}, \orgaddress{\postcode{93106}, \state{CA}, \country{USA}}}

\affil[12]{\orgdiv{Department of Materials Science and Engineering, Stanford University, Stanford}, \orgaddress{\postcode{94305}, \state{CA}, \country{USA}}}
\affil[13]{\orgdiv{Geballe Laboratory for Advanced Materials, Stanford University, Stanford}, \orgaddress{\postcode{94305}, \state{CA}, \country{USA}}}
\affil[14]{\orgdiv{Department of Physics and Center for Quantum Science and Technology, National Tsing Hua University}, \orgaddress{\city{Hsinchu}, \postcode{30013},\country{Taiwan}}}
\affil[15]{\orgdiv{Department of Human and Environmental Studies, Kyoto University}, \orgaddress{\city{Kyoto}, \postcode{606-8501}, \country{Japan}}}


\abstract{ARPES studies have established that the high-$T_c$ cuprates with single and double CuO$_2$ layers evolve from the Mott insulator to the pseudogap state with a Fermi arc, on which the superconducting (SC) gap opens. 
In four- to six-layer cuprates, on the other hand, small hole Fermi pockets are formed in the innermost CuO$_2$ planes,  indicating antiferromagnetism.
Here, we performed ARPES studies on the triple-layer Bi$_2$Sr$_2$Ca$_2$Cu$_3$O$_{10+\delta}$ over a wide doping range, and found that, although the doping level of the inner CuO$_2$ plane was extremely low in underdoped samples, the $d$-wave SC gap was enhanced to the unprecedentedly large value of $\Delta_0\sim$100 meV at the antinode and persisted well above $T_{{c}}$ without the appearance of a Fermi arc, indicating a robust ``nodal metal''. 
We attribute the nodal metallic behavior to the unique local environment of the inner clean CuO$_2$ plane in the triple-layer cuprates, sandwiched by nearly optimally-doped two outer CuO$_2$ planes and hence subject to strong proximity effect from both sides. In the nodal metal, quasiparticle peaks showed electron-hole symmetry, suggesting $d$-wave pairing fluctuations. Thus the proximity effect on the innermost CuO${_2}$ plane is the strongest in the triple-layer cuprates, which explains why the $T_c$ reaches the maximum at the layer number of three in every multi-layer cuprate family. }




\maketitle

\begin{figure}[ht]
\centering
\includegraphics[width=13.5cm]{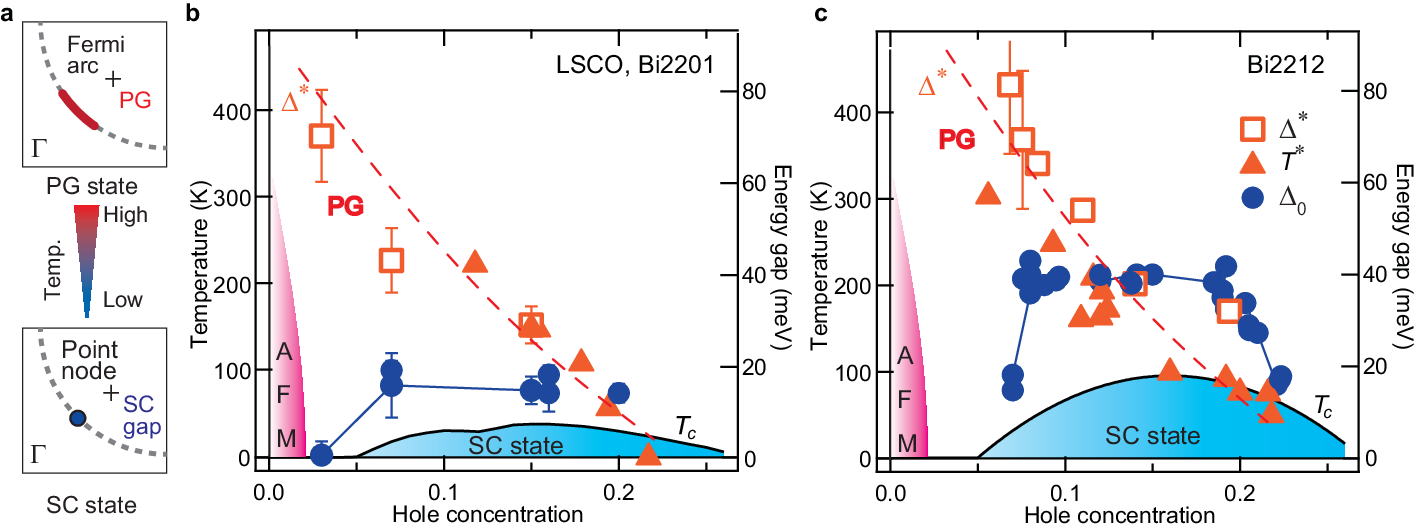} 
\caption{$\bf{a}$ Electronic states on the Fermi surface. Upper panel: Pseudogap (PG) state above $T_{{c}}$, showing a Fermi arc in the nodal region. The Fermi arc is shown by a red curve. Lower panel: SC state below $T_c$, showing a point node of the $d$-wave SC gap. 
Here, the point node is shown by a red circle.
$\bf{b}, \bf{c}$ The phase diagram of single- and double-layer cuprate superconductors~\cite{Yoshida_PRL2009, Kondo_Nature2009, Kondo_PRL2007, Gq_Zheng_PRL2005, JC_Campuzano_PRL1999, KTanaka_Science2006, WSLee_Nature2007, Vishik_pnas2012}. The superconducting (SC) gap extrapolated from the node to the antinode $\Delta_0$, the antinodal pseudogap $\Delta^{*}$, and the pseudogap temperature $T^{*}$ are shown. Grey dashed lines are guide to the eye for $\Delta^\ast$ and $T^\ast$. \label{fig:Fig1}}
\end{figure}

How the superconductivity emerges in the cuprate superconductors through carrier doping into the Mott insulator and what the microscopic origin of the pseudogap have been outstanding issues in condensed matter physics. 
In single- and double-layer cuprates, antiferromagnetism disappears and superconductivity appears at the hole concentration of $\sim$0.05 ~\cite{Tallon_PRB1995}. 
Above the superconducting (SC) critical temperature ($T_{c}$), the $d$-wave SC gap closes near the node while the gap remains open as a pseudogap in the antinodal region up to the ``pseudogap'' temperature $T^{\ast}$ ($>T_c$)[Fig.~\ref{fig:Fig1}(a)]~\cite{Ding_Nature1996, Yoshida_PRL2009}.  
Thus an arc-like Fermi surface, called Fermi arc, is formed in the nodal region in the temperature range $T_{{c}}<T<T^\ast$~\cite{Norman_Nature1998} (Fig. \ref{fig:Fig1}(a)). 
For the single-layer cuprates La$_{1-x}$Sr$_x$CuO$_{4}$ (LSCO), Bi$_2$Sr$_2$CuO$_{6+\delta}$ (Bi2201), and the double-layer cuprates Bi$_2$Sr$_2$CaCu$_2$O$_{8+\delta}$ (Bi2212), the magnitude of the SC gap $\Delta_0$ extrapolated from the node to the antinode, the magnitude of the pseudogap $\Delta^{\ast}$ in the antinodal region, and the temperature $T^{\ast}$ below which the pseudogap opens vary as functions of the hole concentration, as shown in Figs.~\ref{fig:Fig1}(b) and \ref{fig:Fig1}(c).  
$\Delta^{\ast}$ and $T^{\ast}$ are similar between the single- and double-layer cuprates while the SC gap $\Delta_0$ and $T_c$ are very different between them~\cite{Yoshida_PRL2009}. 
The origin of the pseudogap has been extensively studied both experimentally and theoretically, but still under debate as to whether it is a precursor to the superconductivity, namely, pairing fluctuations~\cite{TAO19971507, Bilbro_NatPhys2011, Tallon_PRB2011, Renner_PhysRevLett1998, Yang_Nature2008, Yang_PRL2011, Tu_SciRep2019}, an ordered phase that competes with superconductivity~\cite{Chen_Science2019, Kondo_Nature2009, Uykur_JPSJ2013, Hinkov_Science2008, Sato_NatPhys2017, Ishida_JPSJ2020, Verma_PRB2006, KTanaka_Science2006, Hashimoto_NaturePhys2010}, 
or phenomena arising from strong electron correlations such as electron fractionalization~\cite{Sakai_PRL2013, Sakai_PRL2016, Yamaji_PRR2021, Singh_NatComm2022}.

In the triple-layer cuprate Bi$_{2}$Sr$_{2}$Ca$_{2}$Cu$_{3}$O${_{10+\delta}}$ (Bi2223), the carrier concentration difference between the outer CuO$_2$ plane (OP) and inner CuO$_2$ planes (IP) is large compared to the other multi-layer cuprates families~\cite{Ideta_PRL2010,Ideta_PRL2021,Iwai2014,Luo2023}. 
Indeed, we observed that, in the IP of the underdoped (UD) Bi2223, the doped hole concentration was nearly zero within the accuracy of ARPES measurements and Luttinger sum-rule analysis and that the observed SC gap was unprecedentedly large $\Delta_0\sim$80-100 meV. 
This $d$-wave-like gap did not collapse up to $T_{\rm pair}\sim$100-150 K ($\gg T_c$ $\sim$70-80 K) and the system behaves as a ``nodal metal" with a point node. 
The superconductivity with the large $d$-wave gap and the nodal-metallic pseudogap are likely due to the proximity effect from the adjacent OPs with higher carrier and superfluid densities. 
In the temperature range $T_c<T<T_{\rm pair}$, electron-hole symmetry is observed in the ARPES spectra near the Fermi level ($E_{\rm{F}}$) close to the node, indicating that the nodal metal is a precursor of the $d$-wave SC state. 
Above $T_{\rm pair}$, the node finally collapsed to a Fermi arc and became a conventional pseudogap state. 

So far, there have been several reports on different kinds of ``nodal metals" in the cuprate superconductors. 
Kanigel $et\ al$.~\cite{Kanigel_NatPhys2008} have extrapolated the lengths of the Fermi arcs of various cuprates as a function of $T/T^{\rm{\ast}}$ to $T$ = 0, and concluded that the ground state of the pseudogap state has a Fermi arc of zero length, that is, the ground state is a nodal metal with a point node, but they have not observed the actual point node directly. 
Chaterjee \textit{et al.}~\cite{Chatterjee_NatPhys2010} observed a $d$-wave-like pseudogap with a point node in a lightly-doped \textit{non-SC} cuprate. 
The present finding of the ``nodal metal" above the $T_c$ of the SC samples is qualitatively similar to the persistent point node observed by high-resolution laser ARPES studies of the optimally-doped (OPT) and overdoped (OD) double-layer Bi$_2$Sr$_2$CaCu$_2$O$_{8+\delta}$ (Bi2212) by Kondo \textit{et al.}~\cite{Kondo_NatCommu2015}. However, they have concluded that the point node persists up to $\sim$20-40 K above $T_c$ from the analysis of life-time broadening, and the direct observation of the point node in their raw data has been limited only to $\sim5-10$ K above $T_c$. 
The present observation of a robust $d$-wave SC-like ``nodal metal" in the inner clean CuO$_2$ plane is contrasted with the observation of hole pockets of the antiferromagnetic (AFM) origin in the inner CuO$_2$ planes of 5- to 6-layer cuprates~\cite{Kunisada_Science2020,Kurokawa_NatComu2022}, where superconductivity is weakened compared to the outer CuO$_2$ planes and the CuO$_2$ planes of the singe- and double-layer cuprates. 
The present results should lead us to elucidate the electronic states of the triple-layer high-$T_{{c}}$ cuprates that bring the highest $T_{{c}}$ in every multi-layer cuprate family. Particularly, the observed very large $\Delta_0\sim$100 meV may give us a key for room temperature superconductors.

\section*{Results}
\begin{figure}[ht]
\centering
\includegraphics[width=13.2cm]{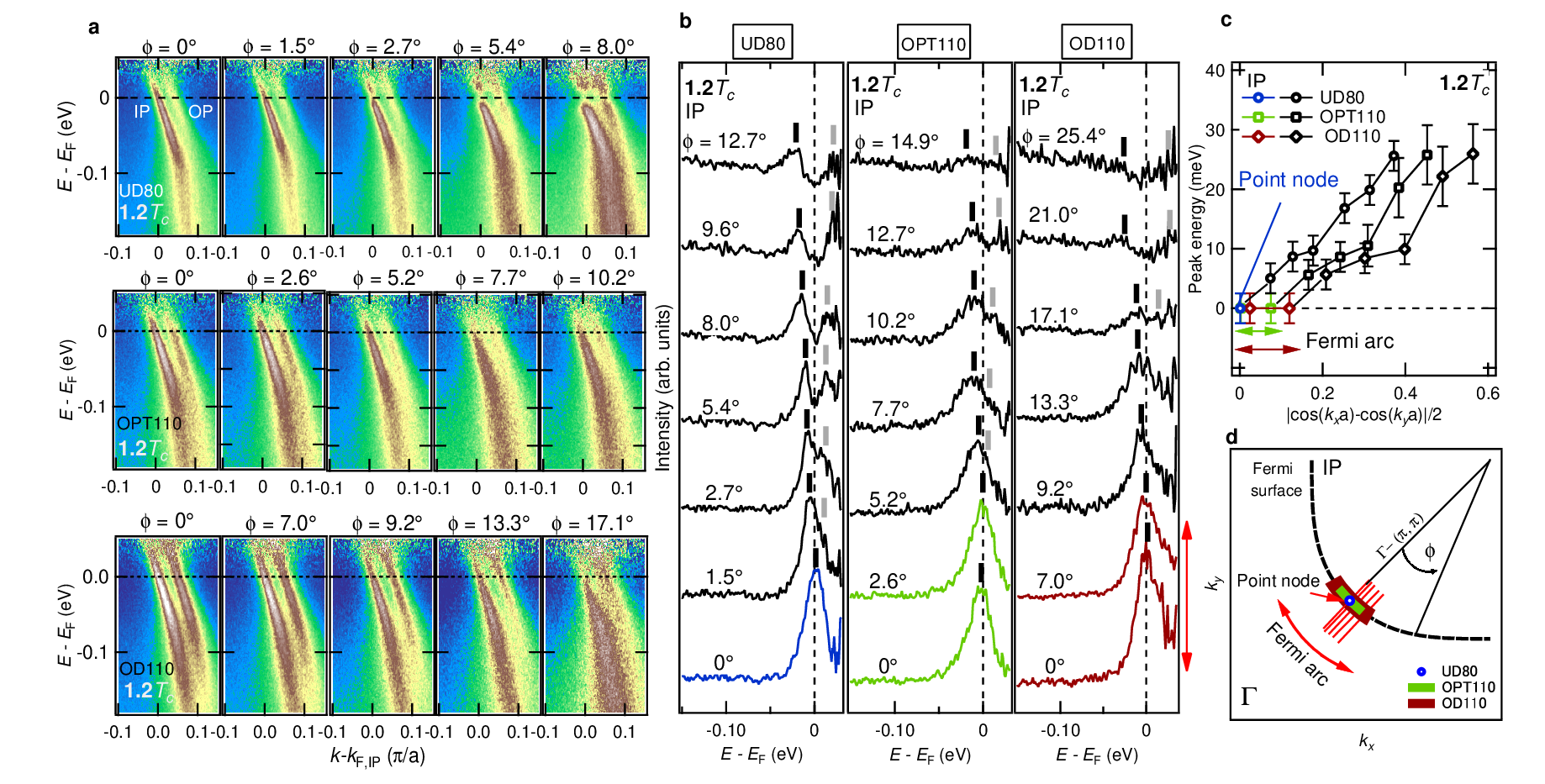}
\caption{ARPES spectra in the nodal region of the normal state ($T = 1.2T_c$) of Bi2223. {\bf a} Intensity $E$-$k$ maps of UD80, OPT110, and OD110 measured for cuts with Fermi-surface angles ($\phi$) defined in panel {\bf d}. Each intensity map has been divided by the Fermi-Dirac (FD) distribution function. {\bf b} EDC divided by the FD function at several $k_{\rm F}$. Black and gray bars indicate peak positions on the occupied and unoccupied sides of $E_{\rm F}$, respectively. EDCs that show a peak at $E_F$ are shown in color.  {\bf c} The momentum dependence of the energy of the black peak for each sample. The Fermi arc for each sample is drawn using the corresponding color. Error bars in {\bf{c}} represent an uncertainty of the spectral peak positions. {\bf{d}} The momentum region where the point node and the Fermi arc exist shown by a circle and a thick line, respectively. See Sec.~\ref{S3} and Fig.~\ref{fig:FigS3} of SI for the Fermi arc lengths of all the samples.}
\label{fig:Fig2}
\end{figure}

\begin{figure}[ht]
\centering
\includegraphics[width=13.8cm]{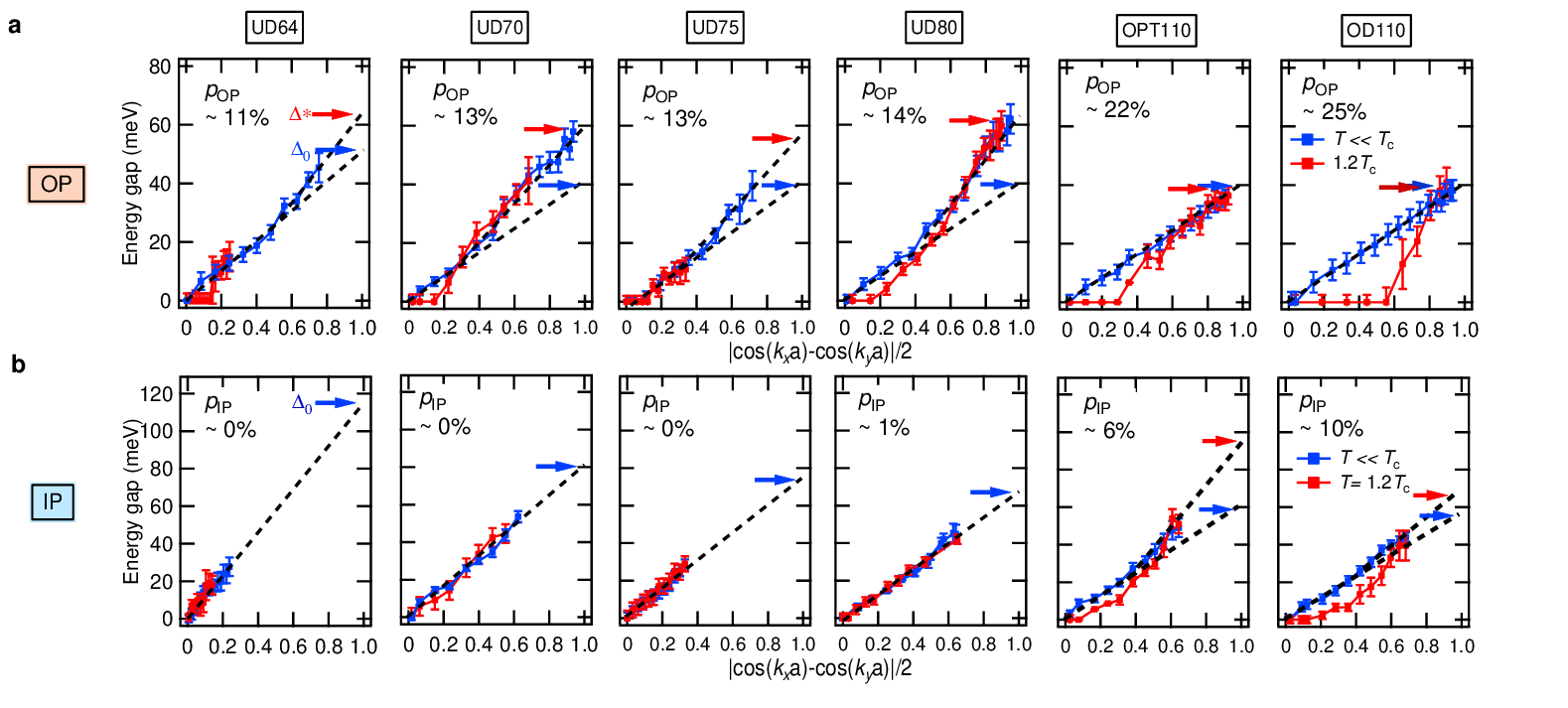}
\caption{Energy gap evolution below and above $T_c$ in Bi2223. {\bf a, b} momentum dependences of the energy gap of the OP and IP bands in the superconducting ($T\ll T_c$) and pseudogap states ($T=1.2T_c$). The nodal gap $\Delta_{0}$ is defined by the gap value extrapolated to the antinode $k=(\pi, 0)$ (blue arrows). In samples showing the ``two-gap'' momentum dependence, the larger energy gap extrapolated to the antinode $\Delta^\ast$ is indicated by red arrows. The IP of all the underdoped samples shows a $d$-wave pseudogap even at $T=1.2T_c$, indicating that $T_{\rm pair}>1.2T_c$. Error bars of energy gap represent standard deviations of the spectral peak positions.}
\label{fig:Fig3}
\end{figure}

\begin{figure}[hbt]
\centering
\includegraphics[width=13cm]{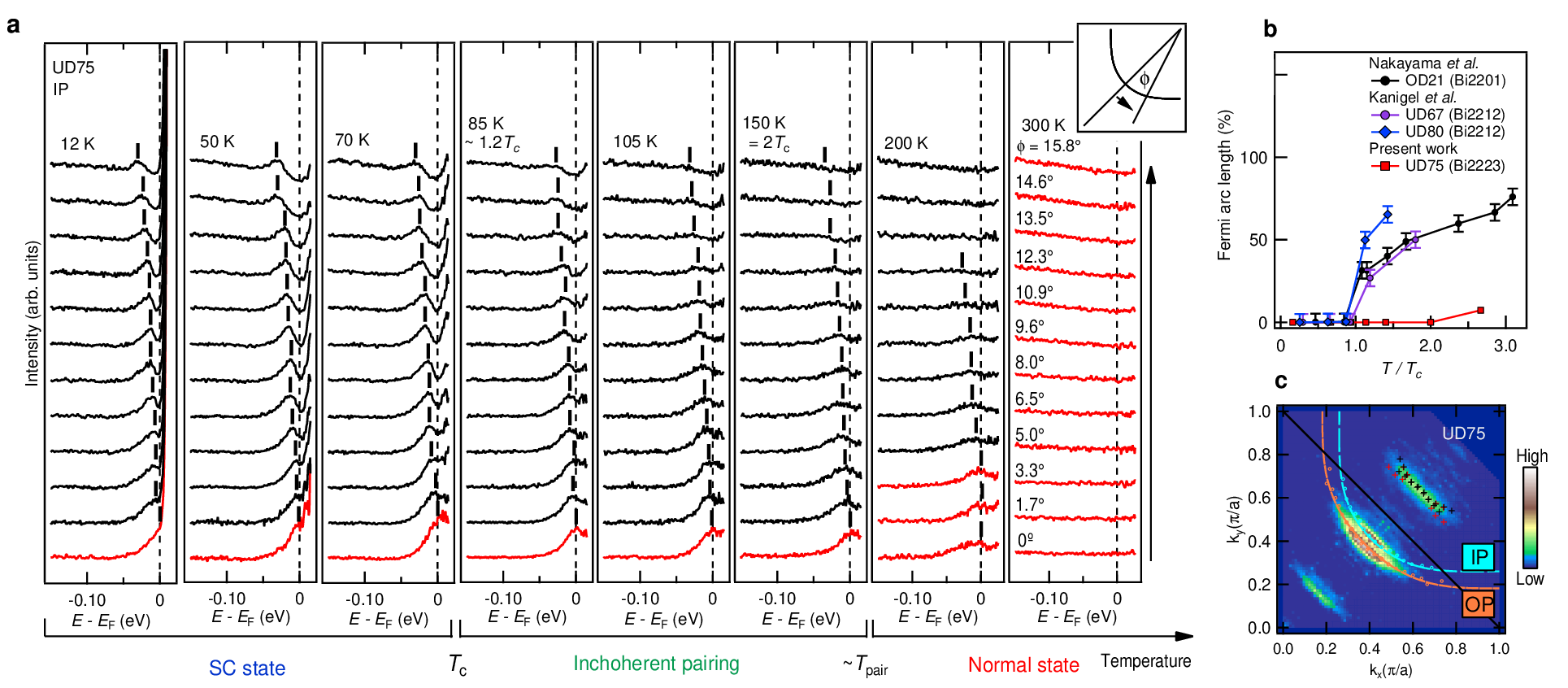}
\caption{Temperature-dependent of the ARPES spectra of the UD75 sample. {\bf a} Temperature dependence of EDCs at $k_{\rm F}$ from $\phi=0^\circ$  to 15.8${^\circ}$ divided by the FD function for the IP band of the UD75 sample. Black bars indicate the peak positions of the EDCs. Red EDCs show a peak at $E_{\rm F}$, indicating that the EDC's are on the Fermi arc. $\bf{b}$ Fermi arc length plotted as a function of $T/T_c$ in single-, double-, and triple-layered cuprates. The data of the single- and triple-layer cuprates are taken from Refs. \cite{Nakayama_PRL2009, Kanigel_PRL2007}. $\bf{c}$ Fermi-surface map of UD75 of Bi2223 at $T$ = 10 K. Red and black markers are replica Fermi surfaces due to the superstructure of the Bi-O layers and those of shadow bands, respectively. }
\label{fig:Fig4}
\end{figure}

\begin{figure}[ht]
\centering
\includegraphics[width=10cm]{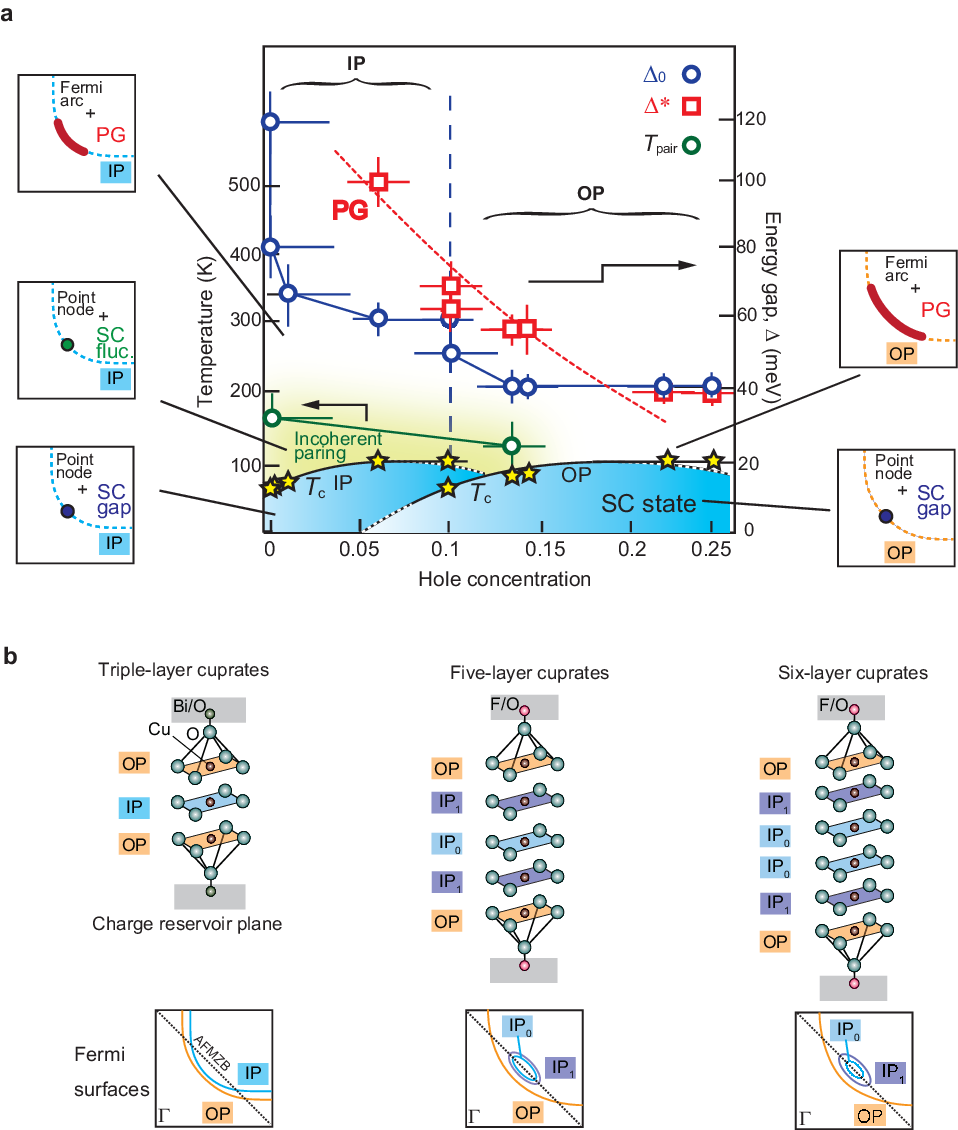}
\caption{Variation of the energy gaps and characteristic temperatures in Bi2223. {\bf{a}} $\Delta_{0}$, $\Delta^{*}$, $T_c$, and $T_{\rm pair}$ of Bi2223 plotted as functions of carrier concentration for each of the OP and IP. Energy gaps $\Delta$ and temperatures $T$ are scaled by $2\Delta=4.3k_{\rm B}T$.The changes in the Fermi surfaces of IP and OP are schematically illustrated for the temperature regions $T < T_c$, $T_c < T < T_{\rm{pair}}$, and $T_{\rm{pair}} < T$.  Uncertainties in the energy gap, temperature, and hole concentration are shown by error bars. {\bf b} Schematic illustration of the crystal structures of multi-layered cuprates. In the 5- and 6-layer cuprates, hole pockets have been observed in the innermost CuO$_2$ plane due to the AFM order \cite{Kurokawa_NatComu2022, Kunisada_Science2020}, whereas the $d$-wave-like pseudogap characteristic of a nodal metal was observed in the triple-layer cuprates.}
\label{fig:Fig5}
\end{figure}

Figure \ref{fig:Fig2} shows the ARPES spectra measured at $T=1.2T_{{c}}$ of the UD80, OPT110, and OD110 samples for cuts shown in panel (d).
[For the samples, see Methods and Sec.~\ref{S1} of Supplementary Information (SI)]
Figure \ref{fig:Fig2}(a) shows the ARPES energy-momentum ($E$-$k$) intensity maps divided by the Fermi-Dirac (FD) function at various Fermi-surface angles ($\phi$)~\cite{Yang_Nature2008, Balatsky_PRB2009, Matsui_PRL2003}, and Fig.~\ref{fig:Fig2}(b) shows the corresponding energy-distribution curves (EDCs) on the Fermi surface ($k = k_{\rm{F}}$) for the IP band. For the OPT110 and OD110 samples, the gap is closed in a finite momentum range around the node ($\phi=0^\circ$), as indicated by black bars in panel (b), and a Fermi arc is formed, as plotted in panel (c). 
For the UD80 sample, the EDC peak is located at $E_{\rm{F}}$ only at the node. 
From the node to the off-node ($\phi>0^\circ$), the peak is shifted towards higher binding energies, like the $d$-wave SC gap, as plotted in Fig.~\ref{fig:Fig2}(c).  
Here, only the IP band of UD80 shows a $d$-wave-like gap at $T = 1.2T_{{c}}$. 
The range of the Fermi arc in momentum space is displayed in Fig.~\ref{fig:Fig2}(d) (see Fig.~\ref{fig:FigS3} in Sec.~\ref{S3} of SI for the other compositions).
Here, we note that a gap with the $d$-wave-like node is possible in the normal state if it is a $d$-density-wave state~\cite{efetov-pepin}. However, such a state induces the spatial modulation of $\textbf{q}=(\pi,\pi)$ that has not been reported in diffraction or scattering experiments on doped cuprates.

Evidence for the $d$-wave pairing fluctuations may be seen in the off-node spectra in Figs.~\ref{fig:Fig2}(a) and \ref{fig:Fig2}(b),  which show that the peaks are electron-hole symmetric in the vicinity of $E_{\rm F}$ (up to $\phi\sim 13^\circ$ for UD80). Here, in UD64-UD80, since the carrier concentration of IP deduced from the Fermi surface area and the Luttinger sum rule  is vanishingly small (Table~\ref{tab1} of SI), AFM order may emerge as in the previous ARPES~\cite{Kunisada_Science2020, Kurokawa_NatComu2022} and NMR~\cite{Kitaoka_PRB1995, KOTEGAWA2001171} studies on (apical-fluorine) multi-layer cuprates. However, even in the ARPES measurement of UD64 at  $T=$ 7 K (not shown), we could not detect AFM band folding, indicating that the N\'{e}el temperature $T_{\rm{N}}$ of IP, if finite, is below 7 K (See Sec.~\ref{S4} of SI). The absence of a hole pocket would, therefore, be attributed to the absence of AFM ordering in the IP of the present UD Bi2223.

Fermi-surface mapping of all the samples (Fig.~\ref{fig:FigS2} of SI) show two bands corresponding to OP and IP. 
The momentum dependences of the energy gaps for the measured samples in the SC ($T\ll T_{{c}}$) and pseudogap ($T$ = 1.2$T_{{c}}$) states are plotted in Figs.~\ref{fig:Fig3}(a) and \ref{fig:Fig3}(b). 
As for the OP [Figs.~\ref{fig:Fig3}(a)], UD64-UD80 shows a ``two-gap'' behavior, OPT110 and OD110 shows a single-component $d$-wave gap, similar to the previous studies on Bi2212~\cite{Vishik_pnas2012}. 
The SC gap $\Delta_0$ (indicated by blue arrows), defined by the extrapolation of the gap from the node to the antinode, is the largest in the most underdoped UD64 ($\Delta_0\sim$ 50 meV), and $\Delta_0\sim$ 40 meV in the other samples. 
The antinodal pseudogap $\Delta^*$ (indicated by red arrows) is the largest in UD64 and UD75 ($\Delta^*\sim$60 meV), but smaller ($\Delta^*\sim$  40 meV) in OPT110 and OD110. 
As for the IP [Fig.~\ref{fig:Fig3}(b)], 
$\Delta_{0}\sim 70-80$ meV for UD70-UD80 and $\sim$120 meV in UD64. 
The latter $\Delta_{0}$ is the largest SC gap reported so far in the high-$T_{{c}}$ cuprates. 
Two-gap behavior is not observed in the UD64-80 samples.

In order to see the temperature range where the point node of IP persists above $T_c$, we measured the ARPES spectra of UD75 over a wide temperature range as shown in 
{Fig.~\ref{fig:Fig4}(a)}. The figure shows EDC's on the Fermi surface ($k = k_{\rm F}$) of the UD75 sample from $T$ = 12 K to 300 K.  Below $T_{{c}}$, the gap closes only at $\phi$ = 0$^\circ$, confirming the $d$-wave SC gap.
The point node persists up to $T\sim$ 150 K, that is,  $T_{\rm pair}\sim$ 150 K, which is twice as high as $T_{{c}}$. At $T$ = 200 K, the point node finally collapsed and a short Fermi arc appeared. At $T$ = 300 K, a well-defined quasiparticle (QP) peak at $E_{\rm F}$ was no longer observed. However, the QP peak was recovered after cooling down to 12 K, indicating that the disappearance of the QP peak at 300 K is not due to the surface degradation or contamination but an intrinsic property of the underdoped IP. See Sec.~\ref{S6} of SI. 
These results suggest that $d$-wave pairing fluctuations or incoherent $d$-wave pairs exist up to temperatures as high as  $\sim2T_c$, that is, $T_{\rm{pair}}\sim 2T_{{c}}$ despite the extremely low carrier concentration of the UD Bi2223. The large pseudogap with the point node in the underdoped Bi2223 may suppress the inter-layer tunneling of the quasiparticle between the planes~\cite{Su_PRB2006}. This explains the extremely strong anisotropy of the quasiparticle mass between the \textit{a-b} plane and the \textit{c} axis in underdoped Bi2223~\cite{Watanabe_PRB2024}.


The temperature dependence of the Fermi arc length of UD75 is plotted as a function of $T/T_c$ in Fig.~\ref{fig:Fig4}(b) and compared with previous studies on the single-layer Bi2201 (OD21)~\cite{Nakayama_PRL2009} and double-layer Bi2212 (UD67, UD80)~\cite{Kanigel_NatPhys2008, Kanigel_PRL2007}.  In Bi2212 and Bi2201, the Fermi arc appeared just above $T_c$, while in the present data on the IP of UD75, no Fermi arc was observed up to $T_{\rm pair}$ which was high as $\sim 2T_c$.  Furthermore, the IP of UD75 did not show a hole pocket of AFM-band folding, in contrast to the 5- and 6-layer apical-fluorine cuprates \cite{Kunisada_Science2020, Kurokawa_NatComu2022} although the hole concentrations were comparably low \ref{fig:Fig4}(c). 
To summarize the present ARPES results, the IP of the underdoped Bi2223 showed a large Fermi surface with a robust node up to well above $T_c$ (nodal metal), and the SC gap and the pseudogap showed the same momentum dependence (see also Sec.~\ref{S7} of SI).

In order to see whether the slightly-doped single CuO$_2$ plane has a tendency to form a nodal metal or not, we have calculated the spectral function $A({\bf k},\omega)$ of the Hubbard model using the Determinant Quantum Monte Carlo method \cite{JARRELL1996133, GunnarssonPRB2010, Moritz_2009, Moritz_PRB2011} (for details, see Sec. ~\ref{S8} of SI). The magnitude of the pseudogap in the antinodal region is about 90 meV at the lowest hole doping of  5$\%$, which is in good agreement with the ARPES experimental results. However, we could not reproduce the nodal metallic behavior, and invariably obtained the formation of Fermi arcs around the $(\pi,\pi)$ direction. This is partly because the spectral width obtained in the calculations remains large in the nodal region, partly because of the high simulation temperatures, but also because a certain amount of hole doping is necessary to induce superconductivity in the Hubbard model~\cite{Jiang_Science2019}. 

\section*{Discussion}

$\Delta_0$, $T_{\rm pair}$, and $\Delta^\ast$ for the OP and IP bands are plotted as a function of hole concentrations in Fig.~\ref{fig:Fig5}(a). 
The figure shows that $\Delta_ 0$, which is constant in the underdoped region in the single- and double-layer systems \cite{Vishik_pnas2012, Yoshida_PRL2009}, increases with decreasing doping in the triple-layer system.
This may be related to the fact that the nodal metal in the triple-layer system is stable compared to the single- and double-layer systems. Besides, the unexpectedly large $\Delta_0\sim$ 80-100 meV is a unique feature of the underdoped clean IP in the triple-layer cuprates that leads to the high $T_{\rm pair}$. If one could fabricate a cuprate superconductor consisting only of CuO$_2$ layers similar to the IP of the underdoped triple-layer cuprates, we would expect to achieve superconductivity at unprecedentedly high temperatures.

As shown in Fig. \ref{fig:Fig5}(a), we found three segmented temperature regions for the IP of the triple-layer cuprates: i) In the SC state ($T < T_c$), the energy gap shows a typical $\it{d}$-wave-like gap with the point node;  ii) Between $T_c < T < T_{\rm pair}$, a pseudogap with a point node (nodal metal) opens due to incoherent pairing (strong SC fluctuations); iii) At $T > T_{\rm pair}$, the energy gap shows a Fermi arc around the node and the pseudogap opens. As for the energy gap of OP, whose hole concentration was above 0.1, a $d$-wave-like gap was observed as IP below $T_c$, but the Fermi arc appeared slight above $T_c$. The greater stability of the nodal metal in the IP of the UD Bi2223 samples (represented by the much larger temperature range $T_c <T< T_{\rm pair}$) than in the single- and double-layer cuprates of similar doping levels may also be due to the proximity of the two OPs with high superfluid densities sandwiching the IP. The proximity effect induces not only the superconductivity in the IP below $T_c$ but also the nodal metallic state in the wide temperature range of $T_{{c}} < T < T_{\rm{pair}}\sim 2T_{{c}}$. It should be noted that, in addition to the reduced disorder as in the case of the IP's of 5- and 6-layer cuprates, the local environment of the IP in the triple-layer UD Bi2223 samples is unique in that it is sandwiched by the nearly optimally-doped OP \textit{on both sides}. The Cooper pairs from both sides may enhance superconductivity in the IP due to the proximity effects, {i.e., pair-tunneling effect. (Single-particle tunneling can also enhance $T_c$, but an unrealistically large value has to be assumed~\cite{Okamoto_PRL2008,Berg-Kivelson}.)} On the other hand, the second-inner plane (IP$_1$) of the 5- and 6-layer cuprates is sandwiched by optimally-doped OP and heavily-underdoped innermost plane (IP$_0$), causing the weaker proximity effect than in Bi2223 as illustrated in Fig.~\ref{fig:Fig5}(b)~\cite{Kunisada_Science2020, Kurokawa_NatComu2022}. 
We consider that these different environments of the inner CuO$_2$ planes leads to their strikingly different behaviors in spite of the similarly extremely underdoped situations,  
namely, the hole pockets resulting from the AFM order in the 5- and 6-layer cuprates against the nodal metallic behavior in the present triple-layer Bi2223, as schematically illustrated in Fig.~\ref{fig:Fig5}(b).
Once the SC state is stabilized in the IP of Bi2223, the clean environment of the IP relatively free from disorder will enhance SC fluctuations above $T_c$ and lead to $T_{\rm pair}$ as high as $\sim2T_c$. 
Note that a pseudogap has been observed up to $\sim1.5T_c$ in another clean system, the Fermi gas of Li atoms~\cite{Li_Nature2023}. 
Recently, the layer-number ($n$) dependence of the $p$-$d$ charge-transfer gap $\Delta$ of the OP in the Bi-based multilayer cuprate series were measured by EELS and was correlated with the $n$ dependence of $T_c$~\cite{Wang-Wang-Zhu2024}. However, the magnitude of the $d$-wave gap $\Delta_0$ of OP is the same between $n=2$ and $n=3$ (Figs.~\ref{fig:Fig1} and \ref{fig:Fig5}), suggesting that the role of IP and/or that of interaction between IP and OP is necessary to explain the highest $T_c$ at $n=3$.

According to the calculation of the CuO$_2$ plane using the extended $t$-$t$'-$t$"-$J$ model~\cite{Shih_LTP2005}, the energies of two possible ground states in the heavily lightly-doped region, that is, the AFM and SC states are nearly degenerate, and are sensitively controlled by the parameter $t$' and $t$" values. Therefore, the AFM and SC phases might be easily switched by an additional parameter. The present experimental results suggest that the SC proximity effect tends to suppress the hole pocket, which arises from AFM order, and to favor the SC state in the triple-layer cuprate, while in the 5- and 6-layer cuprates, the relatively weak proximity effect leaves the electronic states in the innermost CuO$_2$ planes in the AFM ordered phase competing or coexisting with the SC state.
This may explain why the triple-layer cuprates show the highest $T_c$ in each multi-layer cuprate family.

\section*{Methods}
High-quality single crystals of Bi$_2$Sr$_2$Ca$_2$Cu$_3$O$_{10+\delta}$ were grown by the TSFZ method ~\cite{FUJII2001175}. Underdoped Bi2223 samples were successfully obtained in the two-step annealing to control their doping levels ~\cite{ADACHI201553}. Optimally doped and overdoped Bi2223 samples were annealed in oxygen flow and under high oxygen pressure, respectively~\cite{Fujii_PRB2002, FUJII2001175}. Superconducting transition temperature ($T_{c}$) was determined by SQUID measurement and shows $\sim$64, $\sim$70, $\sim$75, $\sim$80, $\sim$110, and $\sim$110 K. They are denoted by UD64, UD70, UD75, UD80, OPT110, and OD110, respectively.

The electronic structure around $E_{\rm F}$ was studied by bulk-sensitive ARPES using low-energy photons. ARPES measurements were carried out at BL7U of UVSOR-III Synchrotron using tunable linearly polarized light of $h\nu$ = 7 eV-18.5 eV photons, and at HiSOR BL-9A using linearly polarized light of $h\nu$ = 7.2 eV and 18 eV. Clean sample surfaces were obtained for the ARPES measurements by cleaving single crystals $in-situ$ in an ultrahigh vacuum better than 6$\times10^{-9}$ Pa. The total energy resolution was at $\sim$5-7 meV. The Fermi level ($E_{\rm{F}}$) of the sample is a reference to that of evaporated gold in electrical contact with the sample. The measurements were performed at well below $T_{{c}}$ ($\sim$10 K) in the SC state and above $T_{{c}}$ (the detailed temperature is shown in the text) in the normal state.


\section*{Acknowledgements}
We would like to acknowledge fruitful discussions with T. Kondo and D. J. Huang. 
ARPES experiments were carried out at UVSOR-III Synchrotron (Proposal Nos. 29-549, 31-572, and 31-861) and at HiSOR (Proposal Nos. 22AG006 and 23BG011). 
This work was supported by the Japan Society for the Promotion of Science (JP20H01861, JP22K03535, JP23K20229, JP23K03317, JP24K06961, and JP25400349) and the National Science and Technology Council of Taiwan (NSTS113-2112-M-007-033). A.F. acknowledges the support from the Yushan Fellow Program under the Ministry of Education (MOE) of Taiwan. C.Y. M. and A.F. acknowledge support from the Center for Quantum Science and Technology within the framework of the Higher Education Sprout Project by the MOE of Taiwan.
The work at Stanford and SLAC (W.O.W., B.M., T.P.D.)  was supported by the U.S. Department of Energy (DOE), Office of Basic Energy Sciences, Division of Materials Sciences and Engineering. Computational work was performed on the Sherlock cluster at Stanford University and on resources of the National Energy Research Scientific Computing Center (NERSC), a Department of Energy Office of Science User Facility, using NERSC award BES-ERCAP0027200.
\section*{Author contributions statement}
S-I.I.,  A.F., and K.T. conceived and coordinated the research guided by physical ideas from T.K.L. and C.Y.M. S.Y., N.S., T.W., S.A., T.N., T.F., S-I.I., and S-I.U. grew the high-quality single crystals of Bi2223. M.A. supported the experiments at HiSOR. W. O.W., B.M., and T.P.D. performed Hubbard-model calculations. S-I.I., T.Y., A.F., T.K.L., and C.Y.M. interpreted the data. S-I.I., T.Y., and A.F. wrote the manuscript with input from all the authors. 

\section*{Competing interests} 
The authors declare no competing financial interests.

\clearpage
\setcounter{page}{1}
\setcounter{table}{0}
\setcounter{figure}{0}
\renewcommand{\thetable}{S\arabic{table}}
\renewcommand{\theequation}{S\arabic{equation}}
\renewcommand{\thefigure}{S\arabic{figure}}
\renewcommand{\thesubsection}{S\arabic{subsection}}
\renewcommand{\thepage}{S\arabic{page}}

\section*{Supplementary Information}
\subsection{Single crystals of Bi2223}
\label{S1}

\begin{figure}[ht]
\centering
\includegraphics[width=10cm]{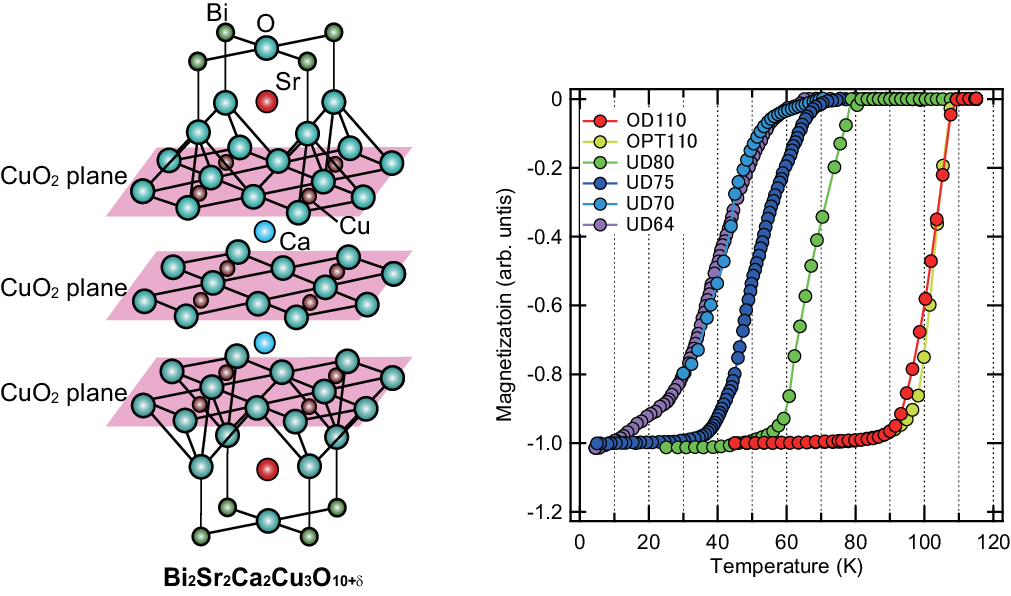}
\caption{Crystal structure of Bi2223 (left panel) and the magnetization of the measured sample (right panel). The triple-layer copper oxide high-temperature superconductor has three CuO$_2$ planes, with the inner CuO$_2$ plane (IP) sandwiched by the outer CuO$_2$ planes (OP). The magnetization is normalized to -1 at the lowest temperature for comparison except for UD70.}
\label{fig:FigS1}
\end{figure}

Figure~\ref{fig:FigS1} (left) shows the crystal structure of Bi2223 investigated in this study. There are three CuO$_2$ planes, and the results of magnetic susceptibility measurements taken at different carrier concentrations are also shown. 
Single crystals of Bi2223 were grown using the traveling-solvent floating-zone (TSFZ) method. Nominal compositions of Bi:Sr:Ca:Cu = (2.25-2.1):(2.0-1.9):2:3 were chosen as the feed-rod compositions. The growth atmosphere was an oxygen-argon mixture with 10$\%$ oxygen, and the growth rate was 0.05 mm/hour. Optimally-doped samples ($T_{{c}}$= 110 K) were synthesized by annealing the crystal in oxygen flow at 500 ${}^\circ$C for 50 hours. Overdoped samples ($T_{{c}} \sim$ 108 - 110 K) were fabricated by annealing them under high-pressure oxygen of 400 atm at 350 ${}^\circ$C for 100 hours. Underdoped samples ($T_{{c}}$ $\sim$ 75 and $\sim$80 K) were synthesized by annealing them at 500 $^\circ$C for 2 hours under an oxygen partial pressure (P$_{\rm{O}_2}$) of 2 Pa and at 600 $^\circ$C for 4 hours under P$_{\rm{O}_2}$ of 0.06 Pa, respectively. Heavily underdoped samples ($T_{{c}}$ $\sim$ 64 and $\sim$70 K) were synthesized commonly by annealing them at 500 $^\circ$C for 12 hours under P$_{\rm{O}_2}$ of 0.6 Pa. In this case, the difference in $T_{{c}}$ is due to a slight difference in their compositions.
The left panel of Fig.~\ref{fig:FigS1} show the measured magnetization of those samples. 
The optimally-doped and over-doped samples showed the same $T_{{c}}$, as previously reported~\cite{Fujii_PRB2002}. The transition in the magnetization curve (right) becomes broader as it is underdoped.

\subsection{Fermi surfaces and hole concentration}
\label{S2}

\begin{figure}[ht]
\centering
\includegraphics[width=10cm]{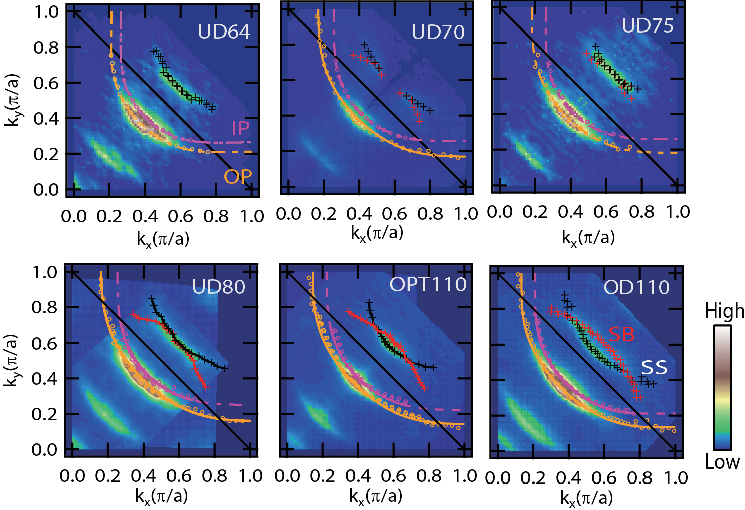}
\caption{Fermi surfaces of Bi2223 for various carrier concentrations. The measurements were made using the photon energy of $h\nu$ = 18.5 eV at low temperatures ($\sim$10 K). 
Replica Fermi surfaces due to the superstructure of the Bi-O plane and shadow band Fermi surfaces are also observed, as indicated by black and red markers, respectively. }
\label{fig:FigS2}
\end{figure}

The Fermi-surface mapping of the measured samples are shown in Fig.~\ref{fig:FigS2} of SI. The photon energy was $h\nu=$ 18.5 eV and measured at low temperatures ($\sim$10 K). In Bi2223 measured in this study, the OP band, which has a large carrier concentration than the IP band, is observed to have a strong spectral intensity. 
Shadow bands (SB) and superlattice signals (SS) due to photoelectron diffraction by the modulated Bi-O planes were also observed. 
The doped hole concentration of each plane has been estimated assuming the Luttinger sum rule and is listed in Table~\ref{tab1}. 

\begin{table}[h]
\caption{Doped hole concentrations of the inner and outer CuO$_2$ planes (IP and OP) of the Bi2223 samples.}
\label{tab1}%
\begin{tabular}{@{}lllllll@{}}
\toprule
CuO$_2$ plane & UD64  & UD70 & UD75 & UD80 & OPT110 & OD110\\
\midrule
OP    & 10±3\%   & 13±3\%  & 13±3\%  & 14±2\% & 22±2\% & 25±2\% \\
IP    & 0±4\%   & 0±4\%  & 0±4\%  & 1±4\% & 6±4\% & 10±4\%\\
\botrule
\end{tabular}
\end{table}

\subsection{Sample dependence of the Fermi arc length}
\label{S3}
\begin{figure}[ht]
\centering
\includegraphics[width=10cm]{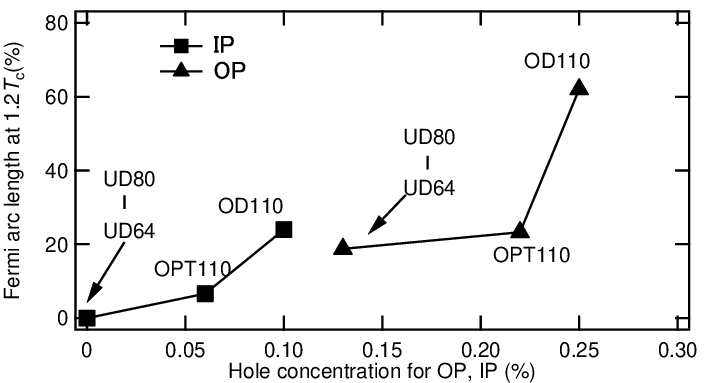}
\caption{Hole concentration dependence of the Fermi arc length in Bi2223 at $T=1.2T_c$}.
\label{fig:FigS3}
\end{figure}

The Fermi arc lengths measured at $T=1.2T_{{c}}$ are plotted in Fig.~\ref{fig:FigS3}. The arc lengths show a tendency to increase with increasing doped hole concentration. The arc lengths of the UD64, UD70, UD75, and UD80 samples are 0 since they show a point node at $T=1.2T_{{c}}$. The uncertainty in the arc length is within the marker size.

\newpage

\subsection{Temperature dependence of ARPES spectra near the nodal region}
\label{S4}
\begin{figure}[hbt]
\centering
\includegraphics[width=13cm]{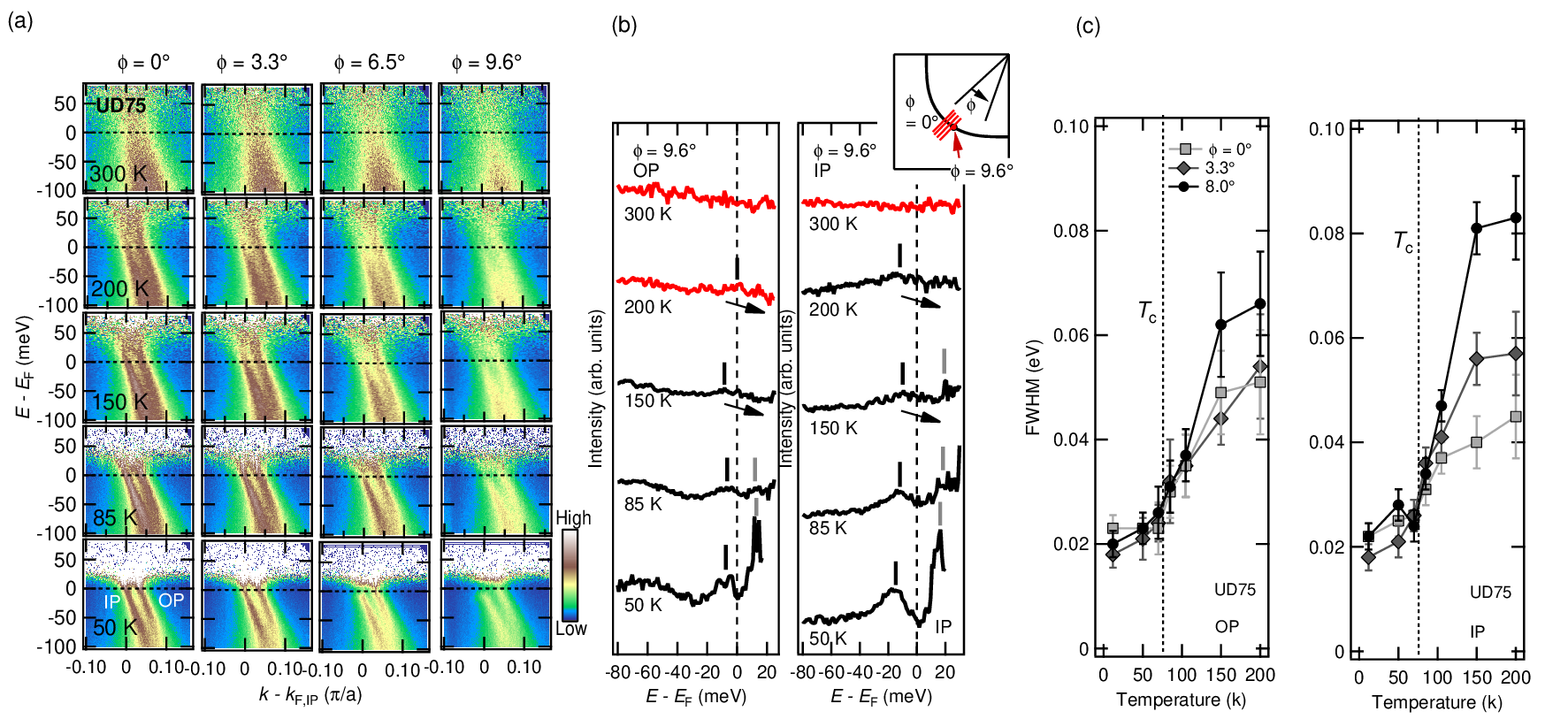}
\caption{Temperature dependence of the ARPES spectra of UD75 around the nodal region. (a) ARPES-intensity map divided by the FD function. The Fermi-surface angle $\phi$ is defined in the inset of (b). (b) Temperature dependence of the EDCs of OP and IP at $k_{\rm F}$ with $\phi=9.6^\circ$. Black bars and gray bars are peak positions in the occupied and unoccupied states, respectively. The black EDCs show a finite gap opening and the red EDCs not. The gap shows electron-hole symmetry up to 105 - 150 K for OP, and up to $\sim$150 K for IP, suggesting that these temperatures can be considered as the $T_{\rm pair}$ of OP and IP. For incoherent electron pairs, the phase strongly fluctuates but the strength of pairing maintains the anisotropy of $d$-wave gap asymmetry. (c) The quasiparticle (QP) width (FWHM) is plotted as a function of temperature near the nodes in the OP and IP bands, respectively. The width of the QP peak begins to decrease from $\sim T_{\rm pair}>T_c$. }
\label{fig:FigS4}
\end{figure}

Figure~\ref{fig:FigS4}(a) shows the temperature dependence of the ARPES-intensity ($E$-$k$) map around the nodal region of UD75. The map at $T$ = 50 K shows that, at finite $\phi$, the electron-hole symmetric gap is open at $E_{\rm{F}}$, indicating the SC state. The EDCs at $k_{\rm{F}}$ of the IP and OP bands at finite $\phi$ (Fig.~\ref{fig:FigS4}(b)) also show electron-hole symmetry for $E_{\rm{F}}$. Above $T_{{c}}$, the intensity above $E_{\rm F}$ is gradually lowered (particularly for IP) and the electron-hole symmetry is gradually lost. This indicates that crossover from the $d$-wave-like nodal metal to the electron-hole asymmetric pseudogap, as reported in the previous studies~\cite{Yang_Nature2008, Sakai_PRL2013} occurs with increasing $T$. The crossover temperature $T_{\rm{pair}}$ depends on the carrier concentration: $T_{\rm{pair}}$ = (2.0-2.7)$T_c$ = 150 - 200 K and 1.1$T_{{c}}$ $\sim$ 85K in the IP band. The temperature dependence of the QP peak width of the ARPES spectra plotted in Fig.~\ref{fig:FigS4}(c) shows a pronounced drop below $T_{\rm{pair}}$. This is consistent with the opening of the $d$-wave-like gap below  $T_{\rm{pair}}$, which suppresses the scattering between QPs.

\subsection{Antiferromagnetic band folding of the IP band in underdoped Bi2223}
\label{S5}
\begin{figure}[hbt]
\centering
\includegraphics[width=10cm]{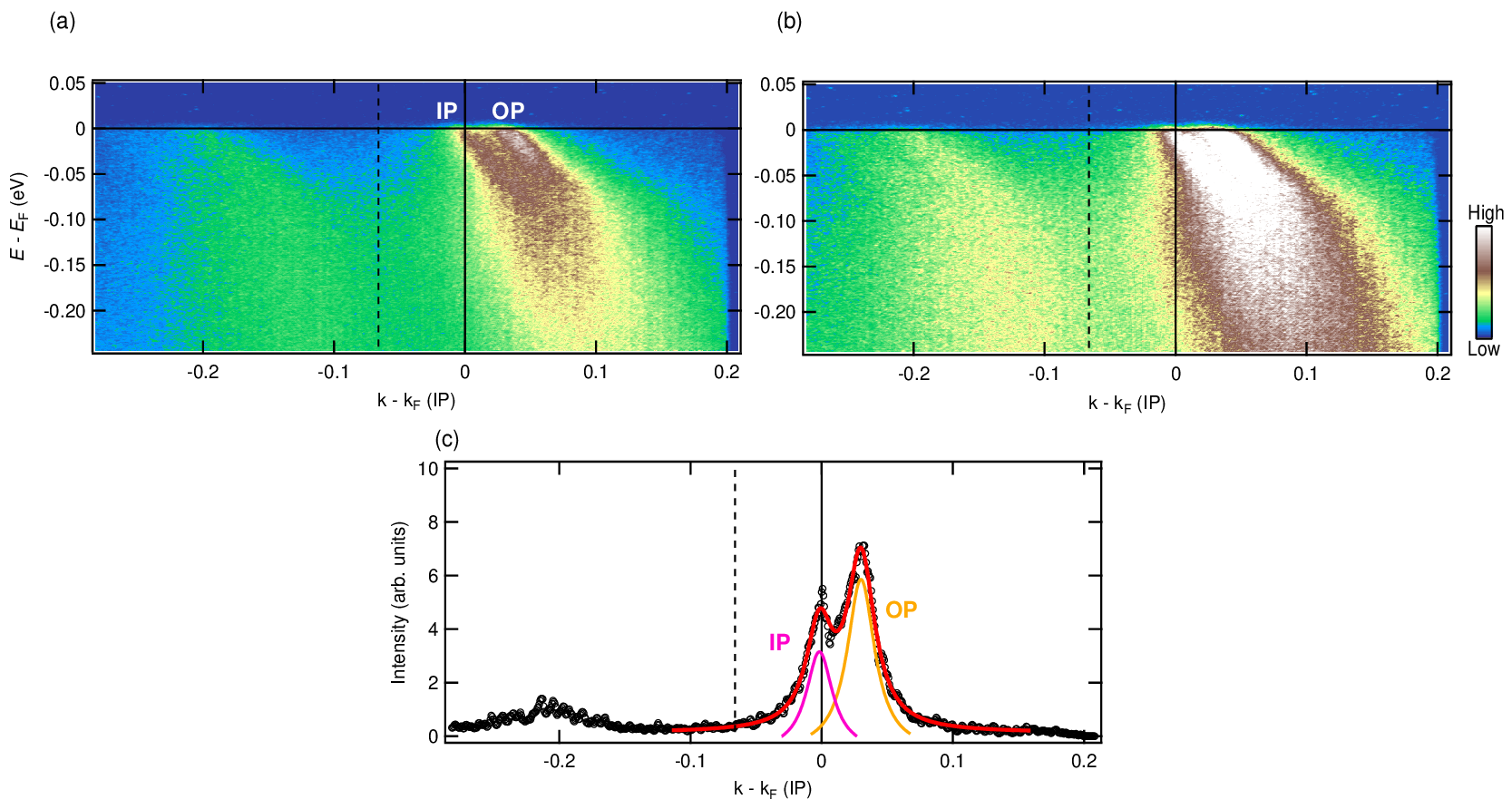}
\caption{ARPES spectra of UD64 measured at $T =$ 7 K. (a) Raw data, (b) Raw data on an intensified intensity scale, (c) Momentum distribution curve at the Fermi level and fitted with two Lorentz functions. A vertical dotted line corresponds to the AFM Brillouin zone boundary. }
\label{fig:FigS5}
\end{figure}

Figure~\ref{fig:FigS5} shows ARPES data of the most underdoped UD64 sample taken at $T$ = 7 K , in order to see whether band folding due to the AFM order was observed or not. It appears that the band due to Brillouin-zone folding was not observed as shown in panel (a). The plot in panel (b) is shown on a different intensity scale, but no band-fold signals are seen. For better visualization, the momentum distribution curve (MDC) is shown in panel (c), fitted with two Lorentzians. If the folded band has been observed, a structure would have been observed in the negative momentum region, but we could not find such a structure.

\newpage

\subsection{Aging effect of ARPES spectra}
\label{S6}
In this study, the temperature is varied from the lowest temperature ($\sim$ 10 K) to room temperature. To confirm the surface contamination, experiments are carried out to check the reproducibility of the data. First, ARPES spectra measured at the low temperature immediately after cleaving are shown in panel (a). The temperature is then increased to 300 K and then cooled to 12 K again. The results are shown in Fig.~\ref{fig:FigS5}(b). Figure ~\ref{fig:FigS5}(c) shows the result of extracting and comparing EDCs at $k_{\rm F}$ of the IP band. It is thus confirmed that there was no change in the line shape and sharpness of the spectrum, indicating that there was no significant contamination or degradation of the sample surfaces during the measurements.

\begin{figure}[ht]
\centering
\includegraphics[width=12.5cm]{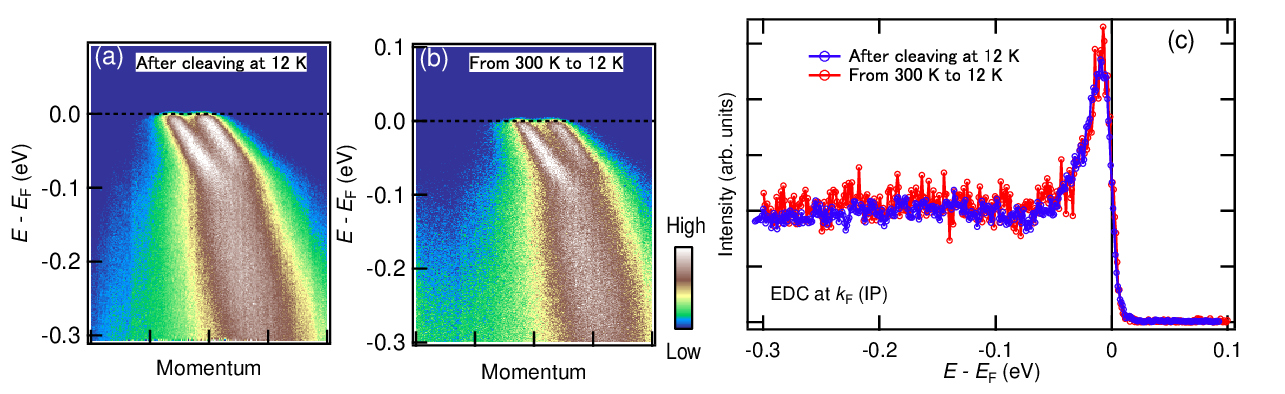}
\caption{Change in ARPES spectra over temperature cycling. (a) ARPES spectrum measured at 12 K after cleavage. (b) ARPES spectrum obtained by increasing the temperature from 12 K to 300 K and then decreasing the temperature to 12 K. (c) Comparison of EDCs obtained at $k_{\rm F}$ of the IP band. The data after the temperature cycling show a spectral line shape similar to after cleavage, although noisier due to a shorter integration time.}
\label{fig:FigS6}
\end{figure}

\subsection{Coherent peak for the inner CuO$_2$ planes of underdoped Bi2223 samples}
\label{S7}
\begin{figure}[ht]
\centering
\includegraphics[width=9cm]{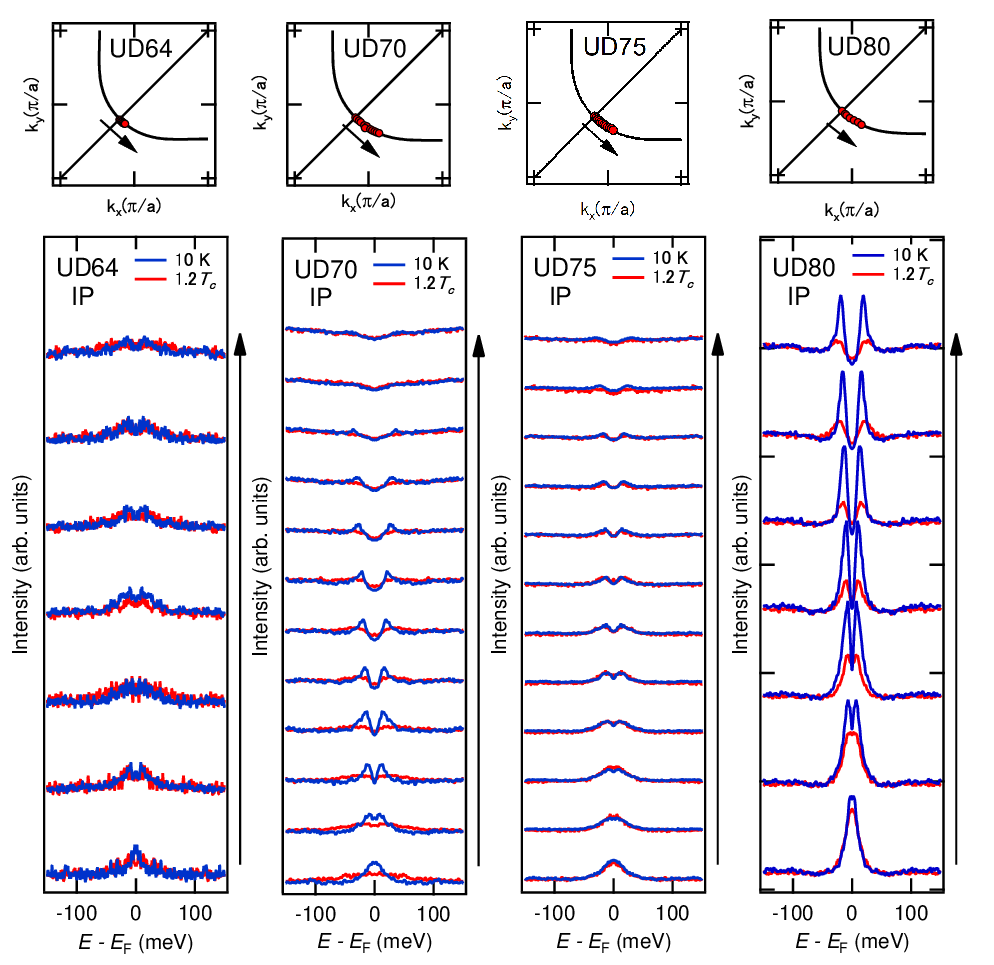}
\caption{Symmetrized EDCs of underdoped Bi2223 samples at the superconducting state (10 K) and pseudogap state (1.2$T_c$). The photon energy is $h\nu$ = 7. 4 eV.  }
\label{fig:FigS7}
\end{figure}

The coherent peak has been observed even in the low carrier concentration for the IP of Bi2223. As shown in Fig.~\ref{fig:FigS7}, we have shown the symmetrized EDCs in the SC and normal states for the sample of UD64-UD80. Even in the UD64 sample, a weak peak is observed, but not so sharp. In going from UD64 to UD80, the intensity of the coherent peaks becomes obvious.

\newpage
\subsection{Carrier-doping dependence of the electronic structure from Hubbard model calculation}
\label{S8}
\begin{figure}[ht]
\centering
\includegraphics[width=13cm]{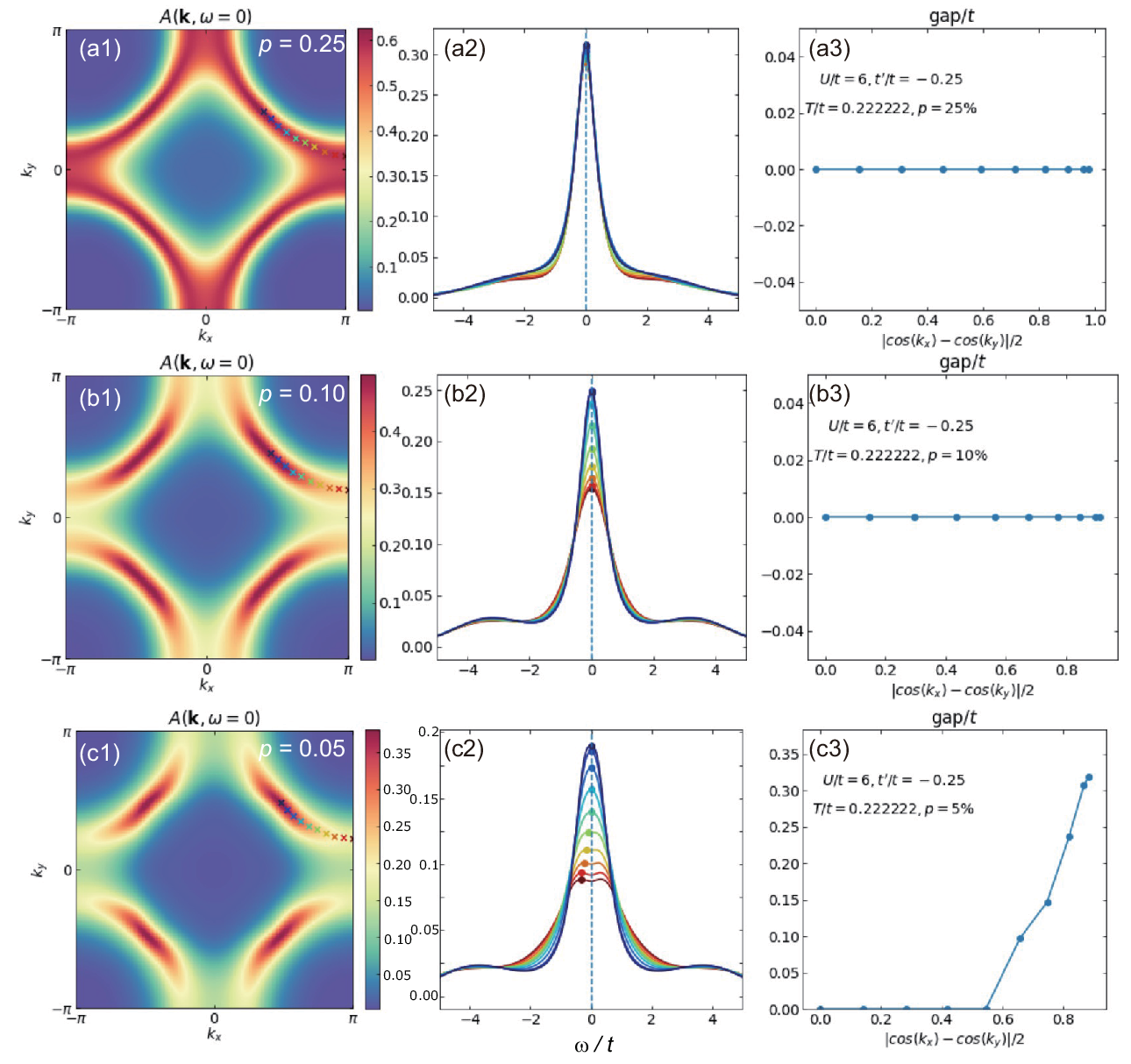}
\caption{Fermi surfaces obtained by Hubbard calculations (left), symmetrized EDC spectra (middle), and momentum dependence of the energy gap (right). In the left panel, $A(\bf{k},\omega=0)$ is the intensity curve representing the Fermi surface, with each symmetrized energy distribution curve for the part of the Fermi surface indicated by the markers in the middle panel, and the energy gap and Fermi arc are shown in the right panel. }
\label{fig:FigS8}
\end{figure}

The Fermi surfaces shown in Figs.~\ref{fig:FigS8}(a1), (b1), and (c1) are calculated for the Hubbard model of the single-layer square lattice for carrier concentrations of 25$\%$, 10$\%$, and 5$\%$, respectively, using the Determinant Quantum Monte Carlo method. Dynamical properties are obtained using the maximum entropy analytic continuation method \cite{JARRELL1996133, GunnarssonPRB2010}. Simulations were performed on 8$\times$8 square lattice clusters with periodic boundary conditions. The overall methodologies are similar to those described in Refs. \cite{Moritz_2009, Moritz_PRB2011}. Here, parameter values used for the calculation are the on-site Coulomb energy $U/t$ = 6, nearest-neighbor hopping $t$ = 0.3 eV, and the next-nearest-neighbor hopping $t^\prime$ with $t^{\prime}/t$ = -0.25 at the temperature of $T/t$ = 0.222 ($\sim$773 K). The EDCs are extracted from the momentum positions indicated by markers on the Fermi surfaces in panels (a1)-(c1). The energy gap is identified by half the energy difference between the peaks of the symmetrized EDC.
 Estimating the size of the Fermi arc and energy gap from the EDC peak positions resulted in the pseudogap opening for the carrier concentration of 5$\%$. The energy of the pseudogap in the antinodal region is $\sim$90 meV, in good agreement with the experiment, but the Fermi arc is found to be longer than the experiment, probably due to the high simulation temperatures and low energy resolution.

\end{document}